\documentclass[9pt,twocolumn,twoside]{optica}
\setboolean{shortarticle}{false}

\title{Precise tuning of single-photon frequency using optical single sideband modulator}

\author[1,*]{Hsin-Pin Lo}
\author[1,**]{Hiroki Takesue}

\affil[1]{NTT Basic Research Laboratories, NTT Corporation, 3-1 Morinosato Wakamiya, Atsugi, Kanagawa, 243-0198, Japan}

\affil[*]{Corresponding author: hsinpin.lo@lab.ntt.co.jp}
\affil[**]{Corresponding author: takesue.hiroki@lab.ntt.co.jp}

\dates{Compiled \today}

\ociscodes{(230.2090) Electro-optical devices; (270.0270) Quantum optics; (270.5585) Quantum information and processing.}

\doi{\url{http://dx.doi.org/10.1364/optica.XX.XXXXXX}}

\begin{abstract}
Frequency translation of single photons while preserving their quantum characteristics is an important technology for flexible networking of photonic quantum communication systems. Here we demonstrate a flexible scheme to interface different-color photons using an optical single sideband (OSSB) modulator. By changing the radio-frequency signal that drives the modulators, we can easily shift and precisely tune the frequency of single photons. Using the OSSB modulator, we successfully erased the frequency distinguishability of non-degenerated photon pairs to obtain the Hong-Ou-Mandel interference with a visibility exceeding $90\%$. We also demonstrated that the level of distinguishability can be precisely controlled by the OSSB modulator. We expect that the OSSB modulator will provide a simple and flexible photonic interface for realizing advanced quantum information systems.
\end{abstract}

\setboolean{displaycopyright}{true}

\begin{document}

\maketitle
\thispagestyle{fancy}
\ifthenelse{\boolean{shortarticle}}{\abscontent}{}
+
Single-photon frequency conversion is an important technology for interfacing the photonic quantum information systems operated at different wavelengths. In the early 1990s, P. Kumar demonstrated the first frequency conversion of non-classical light using a parametric conversion process \cite{Kumar1990, Kumar1992}. This method has been used to demonstrate frequency up- \cite{Tanzilli2005} and down-conversion \cite{Takesue2010} of photons between the visible band and the telecommunication wavelength. In addition, single-photon frequency conversion schemes based on other nonlinear optical interactions such as the four-wave mixing (FWM) \cite{McKinstrie2005,McGuinness2010} and the cross-phase modulation (XPM) \cite{Matsuda2016}, have been demonstrated. The above technologies have also been used to demonstrate photonic quantum interfaces \cite{Takesue2008,Ikuta2011}. 

On the other hand, electro-optic (EO) modulators have been widely used to modulate the amplitude and phase of single or pseudo-single photons in quantum communication experiments, such as quantum key distribution (QKD) based on attenuated laser lights \cite{takesuenp} or single-photon sources \cite{waks}, and they heralded the manipulation of single photons \cite{EOMsinglephoton2008}. Recently, a novel frequency shifter using the EO Doppler shift was also investigated \cite{wright2017}.


In this study, we demonstrate a simpler scheme for single-photon frequency conversion based on an EO device called an optical single sideband (OSSB) modulator \cite{SSB_Izutsu_1981, SSB_Izutsu_2000, SSB_Takesue}. The OSSB modulator can precisely shift the frequency of non-classical photons by using the radio-frequency (RF) signal that drives the phase modulators (PMs). The OSSB modulator can perfectly shift photon frequency to almost only one single sideband. Using the OSSB modulator, we successfully erased the frequency distinguishability of nondegenerate photon pairs by shifting the frequency of one photon. 

\begin{figure}[htbp]
\centering
\fbox{\includegraphics[width=\linewidth]{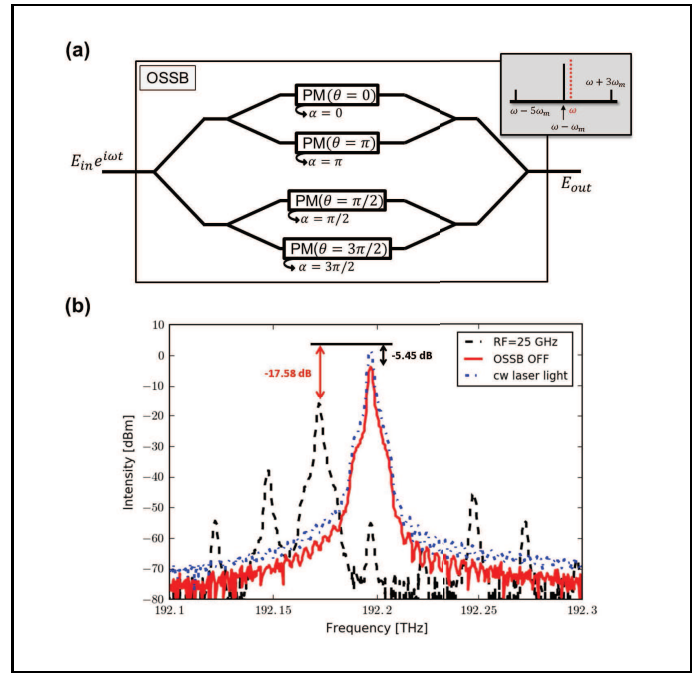}}
\caption{(a) The structure of OSSB modulator with four phase modulators (PMs). The PMs were driven by an RF signal from a synthesizer, and their phase biases were controlled by the DC bias voltages.  The inset shows how the input light with angular frequency $\omega$ is down-converted to the $\omega-\omega_m$ sideband with the OSSB modulator, where $\omega_m$ denotes the angular frequency of the RF signal. (b) Evaluation of the OSSB modulator with classical light. The blue dash-dotted line shows the spectrum of a 1560.9-nm cw light used as the input light for the OSSB modulator. The black dashed and red solid lines are the output spectrums when the OSSB modulator was turned on with a 25-GHz signal and off, respectively. The first sideband (1561.1 nm) is larger than other sidebands by at least 22 dB, which means that most of the input light energy shifts to the first sideband (192.17 THz) when the OSSB modulator is operated.}
\label{ssb}
\end{figure}

The OSSB modulator is composed of integrated lithium niobate (LiNbO$_3$) waveguides, which is based on a nested configuration of four electro-optic PMs \cite{SSB_Izutsu_1981}, as shown in Fig.~\ref{ssb}(a). The input electric field is separated by Y branches and input to two Mach-Zehnder (MZ) modulators. The frequency shift of light is determined by the frequency of the RF signal input into the PMs. The input and output ports of the OSSB modulator both couple with polarization-maintaining fibers.

Here we assume that the input electric field is $E_{in} e^{i\omega t}$, where $\omega$ is the angular frequency of input light, and the output electric field at each phase modulator can be expressed as 
\begin{eqnarray}
&&E_{PM}(\delta, \theta, \alpha) \nonumber \\
&=& \frac{E_{in}}{2} e^{i[\omega t+\delta \sin(\omega_mt+\alpha)+\theta]} \nonumber \\
&=& \frac{E_{in}}{2} e^{i(\omega t+\theta)} \sum_{n=-\infty}^{\infty} J_n(\delta) e^{in(\omega_mt+\alpha)},
\end{eqnarray}
where $J_n$, $\omega_m$, $\delta$, $\alpha$ and $\theta$ denote the $n$th order Bessel function of the first kind, the angular frequency of the modulation signal, the modulation index, the phase bias of the modulation signal at each PM, and the phase bias of the optical carrier at each PM, respectively. 
By setting $\alpha$ and $\theta$ of the four PMs appropriately, the output electric field of the OSSB modulator becomes, for example, \cite{SSB_Takesue}
\begin{eqnarray}
E_{out}&=&\frac{1}{2}\left[ E_{PM} (\delta,0, 0) + E_{PM} (\delta,\pi,\pi) \right. \nonumber \\
&& \left. + E_{PM} (\delta, \pi/2, \pi/2) + E_{PM}(\delta, 3\pi/2, 3\pi/2) \right] \nonumber \\
&=&E_{in} e^{i \omega t} \left[\cdots + J_{-1}(\delta) e^{-i \omega_m t} + J_3 (\delta) e^{i 3 \omega_m t} \right. \nonumber \\
&& \left. + J_{-5} (\delta) e^{-i 5 \omega_m t}+ \cdots \right].
\end{eqnarray}
If we set $\delta$ appropriately, the OSSB modulator functions approximately as a frequency shifter with a frequency shift of $-f_m = -\omega_m/2 \pi$. From this example, it is clear that we can easily shift the frequency downward as shown in the inset of Fig.~\ref{ssb}(a). The theoretical maximum frequency conversion efficiency of the OSSB modulator from the input light to the first sideband ($J_{-1}$) is 33.81 \% and all the other sidebands are negligible ($J_{-3}$ is 0.98 \%). The range of the frequency shift, $\omega_m$, is limited by the bandwidth of the PM, which can reach up to $\sim$ 100 GHz \cite{maxrfpower}. 

We experimentally evaluated the characteristics of the OSSB modulator using classical light. The result is shown in Fig.~\ref{ssb} (b). We input 1560.9-nm continuous-wave (cw) light  whose spectrum is shown by the blue dash-dotted line. The spectrum at the output of the OSSB modulator was obtained as the red solid line when the modulator was turned off. When the OSSB modulator was turned on with a 25-GHz signal, it showed a frequency shift of -25 GHz as shown by the black dashed line. The intensity of the first modulated sideband is 22-dB larger than the others, implying that the effects of the residual sidebands were negligibly small. This confirmed that the OSSB modulator can work as a frequency shifter of cw laser light. The total optical insertion loss as a frequency shifter was $-17.6$ dB, which includes the waveguide propagation loss, and the connection loss between the waveguide and fiber ($\sim$ 5.5 dB). 
The experimental frequency conversion efficiency of the OSSB modulator (-12.1 dB) is smaller than the theoretical maximum efficiency because the limited RF signal power in our experimental setup. 

\begin{figure}[htbp]
\centering
\fbox{\includegraphics[width=\linewidth]{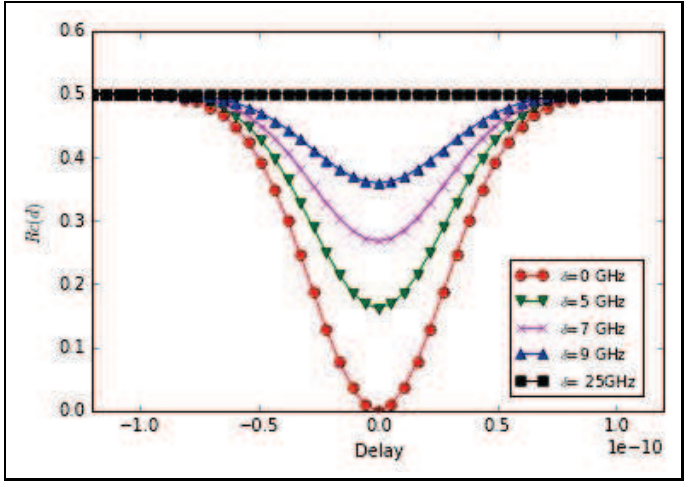}}
\caption{The theoretical plot of Eq. (\ref{3}) with the experimental bandwidth of signal and idler photons. Circles, reversed triangles, x symbols, triangles, and squares show the case when the frequency difference $\delta$ was set at 0, 5, 7, 9, and 25 GHz, respectively, with the HOM visibilities of 1, 0.677, 0.465, 0.282 and 0.}
\label{hom_theory}
\end{figure}

The indistinguishable photon pairs would show a perfect Hong-Ou-Mandel (HOM) dip when they spatially and temporally overlap in a beamsplitter \cite{hom, Lobeamlike2011, Lo2x22011}. When photons with slightly different central frequencies are input into a beamsplitter, HOM visibility will generally decrease. When the central frequency difference is $\delta$, the coincidence rate between signal and idler photons at the output of the beamsplitter is given by \cite{Jin2011, Mosleythesis}
\begin{eqnarray}
R_c(d)=\frac{1}{2}-\frac{{\sigma_s}{\sigma_i}}{\sigma_s^2+\sigma_i^2}Exp\left[-\frac{\sigma_s^2\sigma_i^2 d^2+4\delta^2}{2(\sigma_s^2+\sigma_i^2)}\right], \label{3}
\end{eqnarray}
\label{coinc}
where $\sigma_s$ and $\sigma_i$ are the bandwidths for the signal and idler photons, respectively, and $d$ is the optical delay between them. The visibility of HOM interference with frequency detuned photons is obtained as 
\begin{eqnarray}
V\equiv\frac{R_c(\infty)-R_c(0)}{R_c(\infty)}=2\left(\frac{{\sigma_s}{\sigma_i}}{\sigma_s^2+\sigma_i^2}\right)Exp\left[\frac{-4\delta^2}{2(\sigma_s^2+\sigma_i^2)}\right]. \label{4}
\end{eqnarray}
As shown in the original paper on HOM interference, a photon pair exhibits perfect quantum interference when their central wavelength is perfectly overlapped ($\delta$=0). Fig.~\ref{hom_theory} shows the theoretical expectation for the HOM dip obtained by using Eq. (\ref{3}) with a 0-, 5-, 7-, 9- and 25-GHz frequency difference between the signal and idler photons. The HOM visibilities were calculated to be 1, 0.677, 0.465, 0.282, and 0, respectively. The accidental coincidences caused by multi-photon emission or other optical and electrical noise would also decrease the visibility.

\begin{figure}[htbp]
\centering
\fbox{\includegraphics[width=\linewidth]{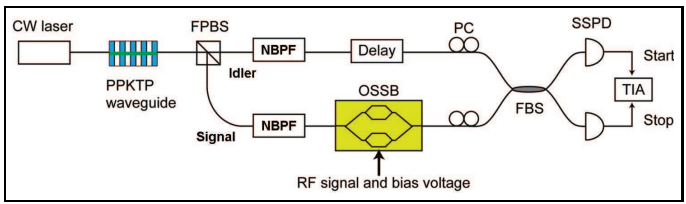}}
\caption{Experimental setup. All the components are connected by fibers. PPKTP: periodically poled potassium titanyl phosphate waveguide, FPBS: fiber polarized beam splitter, NBPF: narrow band pass filter, DL: delay line, OSSB: optical single sideband modulator, PC: polarization controller, FBS: fiber beamsplitter, SSPD: superconductor single-photon detector, TIA: time interval analyzer.}
\label{hom_a}
\end{figure}

\begin{figure}[htbp]
\centering
\fbox{\includegraphics[width=\linewidth]{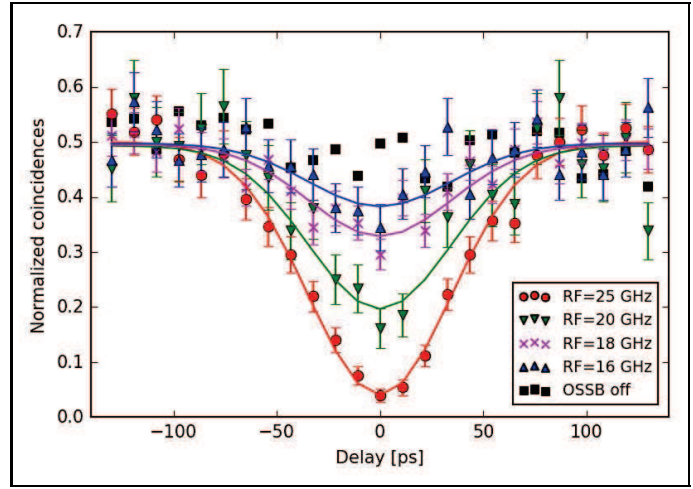}}
\caption{The HOM interference experiment results. The vertical and horizontal axes show the normalized coincidence count and the relative optical delay between the signal and idler photons at the fiber beamsplitter (FBS), respectively. 
Circles, reversed triangles, x symbols, and triangles show the experimental results when the RF was set at 25, 20, 18, and 16 GHz, respectively. The squares shows the case when the OSSB modulator was turned off. The solid lines are the best-fit curves, whose visibilities are 0.920 $\pm$ 0.059, 0.602 $\pm$ 0.042, 0.337 $\pm$ 0.028 and 0.231 $\pm$ 0.038 for the RF of 25, 20, 18 and 16 GHz, respectively.}
\label{hom_b}
\end{figure}

We performed experiments to demonstrate frequency manipulation of single photons using the OSSB modulator. The experimental setup is shown in Fig.~\ref{hom_a}. A PPKTP waveguide was pumped by a cw laser light (central wavelength of 780.5 nm and laser light power of 2 mW) to generate correlated photon pairs via a type-II spontaneous parametric down-conversion (SPDC) process. The signal and idler photons were separated by a fiber polarization beam splitter (FPBS) and then inserted into the narrow band pass filters (NBPF) whose transmission spectra could be well approximated with a Gaussian function with a 3-dB bandwidth of $\sim$10 GHz. The center wavelengths of the NBPFs for the signal and idler were set to minus and plus 12.5 GHz from the half of the pump light frequency, respectively, so that we could prepare two single photons with the same temporal and spatial modes but almost no spectral overlap. 

Then, the signal photon was input into an optical delay line to adjust its temporal position, while the idler photon was launched into the OSSB modulator, which shifted the frequency of the idler photon by -25 GHz. The signal and idler photons were input into a fiber beamsplitter (FBS) followed by two superconducting single-photon detectors (SSPD) for the HOM interference measurement. The polarizations of the signal and idler photons were controlled by the polarization controller (PC), and the polarization of a single photon was filtered by the fiber polarizer in front of the FBS, which is not shown in Fig~\ref{hom_a} for conciseness. The detection signals were used as the start and stop signals for a time-interval-analyzer (TIA), and the coincidences between the signal and idler photons was counted. The coincidence count rate is 1 Hz outside the HOM dip, which includes accidental coincidences. By changing the delay for the signal photon, we can change the temporal overlap between two photons with 10-ps resolution. The coincidence-to-accidental ratio (CAR) \cite{Takesue_CAR, Takesue_CAR_theory} of the photon pair source was approximately 70 under the continuous-wave light pump. 

Figure~\ref{hom_b} shows the normalized coincidences as a function of the relative delay between the signal and idler photons and the best-fitting curves when the OSSB modulator was operated. From Eq. (\ref{3}), the HOM dip vanishes when the frequency difference of  the photon pair is 25 GHz. When the OSSB modulator was turned off (black squares in Fig.~\ref{hom_b}), we did not observe a notable dip in the coincidence counts, which implies that the initial input photons were distinguishable. 

On the other hand, when we turned on the OSSB modulator with a 25-GHz RF signal, we observed a clear HOM dip (red circles) in Fig.~\ref{hom_b}. The visibility of the HOM interference is 0.920 $\pm$ 0.059 without subtracting any noise counts, including accident coincidence ones. Note that we did not place any additional filters after the OSSB modulator to suppress residual sidebands. This result clearly shows that the frequency distinguishability between the photons was erased almost perfectly with the OSSB modulator. 

Figure~\ref{hom_b} also shows the HOM interference results for different frequency shifts. When the RF was set at 16, 18, and 20 GHz,  the HOM visibilities were 0.231 $\pm$ 0.038, 0.337 $\pm$ 0.028, and  0.602 $\pm$ 0.042, respectively. This confirms that the frequency distinguishability was continuously tuned by the proposed scheme. With this scheme, we can flexibly shift the single-photon frequency by accurately tuning the RF signal.

These results are slightly worse than the theoretical prediction shown in Fig.~\ref{hom_theory}. One reason for the difference is the accidental coincidences caused by multipair emission from the photon pair source and the dark count. If we consider the case where the noise photons are included, the theoretical visibility is 0.97 when the frequency distinguishability is completely eliminated\cite{noise}. 
Another possible source of degradation is the loss of the OSSB modulator which reduced the visibility of HOM interference by about 0.4\% (see Supplementary Material 1 for details). A slight misalignment of the NBPF center frequencies can also decrease the visibility. Since we adjusted the center frequencies with an optical spectrum analyzer with a 1-GHz resolution, the frequency deviation of the NBPF would be at most $\sim$ 2 GHz. According to Eq. (\ref{4}), such a frequency deviation results in up to $\sim$4\% visibility degradation. Finally, the effect of the unsuppressed sidebands should also be considered. Based on the formulas shown in Supplementary Material 2, the visibility degradation due to residual sidebands is estimated to be negligibly small with the present experimental condition. With all these factors considered, the HOM visibility is expected to be $\sim$93\%, which agrees well with our experimental result.

In the present experiment, the total conversion efficiency is relatively small, because of the insertion loss of the device and the insufficient RF amplitude. Although there research has sought to decrease the loss of waveguide devices (for example, see \cite{lowlossmodulator1982}), low-loss devices are still hard to obtain on the market.  With the recent motivation for integrated quantum photonics \cite{politi,bonneau}, we can expect that more effort will be made to reduce the losses of waveguide devices at the production level. 
In order to obtain the maximum conversion efficiency, we need to drive the PMs in an OSSB modulator with a modulation index of $\sim$1.84. In our experimental setup, the maximum RF amplitude at 25 GHz was limited to 0.46 V, which corresponds to a modulation index of 0.54. Therefore, the RF amplitude required to obtain the maximum conversion efficiency is estimated to be about 1.6 V, which will be obtainable if we improve the frequency response of our electronics and RF amplifiers with higher power. 

We expect that our scheme may find applications in quantum information systems that require relatively fewer numbers of qubits, such as quantum communication assisted by quantum memories. For example, our scheme will be useful for a measurement-device-independent QKD system with quantum memories, with which the key generation rate scales with the loss over half of the total distance \cite{Abruzzo2014,Panayi2014}. Our scheme might also be useful for quantum communication systems with optical cavities containing matter qubits \cite{Munro2010}.  In such systems, the merit of precise tuning of the single-photon frequency for narrowband quantum memories might outweigh the drawback, namely, the additional loss due to the non-unity conversion efficiency. According to the theoretical investigation in \cite{Sangouard2011}, even with additional loss originating from the frequency converters for each link, we can construct quantum repeater systems that can beat the direct transmission of photonic qubits. 
Another possible way to tune the frequency of a single photon with an accuracy of the RF frequency is to use an acousto-optic (AO) modulator, which was demonstrated to shift the frequency of a single photon by 76 MHz to obtain a HOM interference with photons with slightly different frequencies \cite{leong}. Compared with the frequency tuning using the AO modulator, the OSSB modulator has an advantage in terms of tuning range: while the frequency shift by an AO modulator is limited to at most several hundreds of megahertz, the tuning range of an OSSB modulator will reach up to several tens of gigahertz. Thus, the OSSB modulator will provide simple, flexible and precise tuning of single photons, which is difficult to realize with other existing technologies. Therefore, our scheme will complement existing frequency converters based on nonlinear optics, and possibly be suitable for interfacing photons with telecom-band quantum memories \cite{Saglamyurek}. 

In conclusion, we developed a simple scheme to convert the frequency of single photons using an OSSB modulator and used it to realize the erasure of the frequency distinguishability of single photons. We also demonstrated the flexible tuning of the frequency distinguishability between single photons by changing the frequency of the RF signal that drives the modulator. We expect that this would be a very important technology for interfacing photonic quantum information systems operated at different frequencies. 


The authors thank K. Azuma for fruitful discussions.

\bibliography{sample}

\begin{thebibliography}{1}
\bibitem{noise} The HOM visibility with accidental coincidence is obtained as function of CAR as V$\simeq \frac{CAR-1}{CAR+1}$.
\end{thebibliography}

\ifthenelse{\boolean{shortarticle}}{%
\clearpage
\bibliographyfullrefs{sample}
}{}


\end{document}